\documentclass{article}

\usepackage{arxiv}
\usepackage{cite}

\usepackage[utf8]{inputenc} 
\usepackage[T1]{fontenc}    
\usepackage{hyperref}       
\usepackage{url}            
\usepackage{booktabs}       
\usepackage{amsfonts}       
\usepackage{nicefrac}       
\usepackage{microtype}      
\usepackage{lipsum}
\usepackage{graphicx}
\usepackage{amsmath}
\usepackage{geometry}

\title{Thermal Control of Laser Powder Bed Fusion using Deep Reinforcement Learning}

\author{
  Francis Ogoke \\
  Department of Mechanical Engineering\\
  Carnegie Mellon University\\
  Pittsburgh, PA 15213 \\
  
   \And
 Amir  Barati Farimani \\
  Department of Mechanical Engineering\\
  Carnegie Mellon University\\
  Pittsburgh, PA 15213 \\
 
}

\begin{document}
\maketitle

\begin{abstract}
Powder-based additive manufacturing techniques provide tools to construct intricate structures that are difficult to manufacture using conventional methods. In Laser Powder Bed Fusion, components are built by selectively melting specific areas of the powder bed, to form the two-dimensional cross section of the specific part. However, the high occurrence of defects impact the adoption of this method for precision applications. Therefore, a control policy for dynamically altering process parameters to avoid phenomena that lead to defect occurrences is necessary. A Deep Reinforcement Learning (DRL) framework that derives a versatile control strategy for minimizing the likelihood of these defects is presented. The generated control policy alters the velocity of the laser during the melting process to ensure the consistency of the melt pool and reduce overheating in the generated product. The control policy is trained and validated on efficient simulations of the continuum temperature distribution of the powder bed layer under various laser trajectories.

\end{abstract}


\section{Introduction}

Additive Manufacturing (AM) enables the construction of three-dimensional parts that are difficult to make using traditional machining methods. These methods enable complex part production while decreasing material waste, and allow for local control of materials and mass customization \cite{markl2016multiscale}.   Laser Powder Bed Fusion is a subcategory of AM, which creates a molten product by using a heat source to fuse layers of metallic powder together. Powder Bed Fusion methods have been used to construct complex lattice products from alloys of iron and titanium, and have seen heavy usage in the biomedical and aerospace industries. However, challenges lie in the widespread adoption of these methods, due to the tendency for PBF-produced parts to develop defects and inferior physical properties due to residual stresses that may lead to failure in these specific applications \cite{cunningham2017analyzing,mower2016mechanical,spierings2013fatigue,lewandowski2016metal}.These defects include poor surface finish, increased porosity, delamination, cracking, and residual stresses, leading to inferior mechanical properties and poor geometric conformity.

Previous experimental studies have shown that the characteristics of the molten area associated with the scanning process are significant contributors to the formation of defects in the finished product. The melt pool can produce keyhole porosities, while the temperature gradients produced during melting can also influence the microstructure that forms and lead to cracks \cite{mani2015measurement,murr2012metal,al2010origin}. To avoid defects due to adverse melt pool behavior during the scan path as well as overheating, one would ideally be able to adjust the process parameters to the changing temperature distribution during the scan trajectory. Powder Bed Fusion is an inherently complex multiscale process, as physical effects that occur at both the powder and continuum scales determine the properties of the final material. This work will focus on effects at the continuum scale, neglecting convection and radiation heat transfer of the heat source to consider on the effect of heat conduction on the resulting temperature field.   This abstraction creates the necessary speed-ups required to run relatively data-intensive DRL algorithms.

 Classical optimization methods have been used in the past to produce control strategies to reduce the occurrence of mechanical defects. However, these methods require severe model order reductions, and are restricted in the amount of data that they are able to process due to computational expense. Evolutionary algorithms, such as deep reinforcement learning, show promise in adapting to previously unseen process parameters and dynamically changing conditions. Purely statistical methods have also been used to model the AM process, such as analysis of variance, ANN, and response surface methodology, though these data driven approaches are restricted due to a lack of physical insight.
Previous work has been done applying classical control strategies to dynamically change process parameters during the melting process \cite{Mohanty2013,Yeung2019,Papacharalampopoulos2018,Craeghs2010,Lee2016,Afazov2017a,Afazov2017,Druzgalski2020,Clijsters2014,Ganeriwala2019,Wang2020}. In \cite{renken2018model}, a model-based feedforward closed-loop control method is presented to limit temperature deviations, using an FPGA based controller. In \cite{Wang2020}, an open-loop control mechanism for limiting the deviation of the melt-pool area during the melting process is developed using PID methods. However, the use of PID controllers requires extensive parameter tuning, and open-loop control is unable to dynamically change the process parameters in response to feedback from the system. Deep Reinforcement Learning methods are able to automatically learn a control policy that minimizes the error signal with minimal manual tuning required.  Previous attempts of addressing the instability in the melt pool depth have used classical optimization techniques, such as greedy heuristics, but rely on model reductions, with assumptions that may not be satisfied with arbitrary scan paths \cite{forslund2019greedy}. In response to these challenges, data-driven optimization techniques have seen recent adoption in the field of Additive Manufacturing \cite{Li2018,Schwalbach2019,Tapia2018 }.  \cite{yang2019investigation} used Convolutional Neural Networks to predict characteristics of the melt pool depth given an image of the melt pool itself, while \cite{liu2021physics} uses Physics Informed Machine Learning to analyze the mechanics of pore generation.

 Deep Reinforcement Learning (DRL) has emerged in recent years as a method of addressing complex control scenarios in relatively high-dimensional spaces\cite{verma2018efficient,beintema2020controlling,rabault2019accelerating,ren2020applying}.  DRL is a semi-supervised learning method that iteratively improves on an initially random control policy by collecting experience and feedback from the simulated environment. DRL-based methods have been successful in transport phenomena and process control applications, such as the control of self-propelled swimmers\cite{verma2018efficient}, and the control of Rayleigh-Bernard Convection \cite{beintema2020controlling}.  These advances show promise in its ability to perform optimization for complex, high dimensional state spaces. Here, we present a DRL framework to create a sophisticated control policy that addresses a key mechanism of defect formation, the variation of the melt pool dimensions during the melting process, with minimal manual parameter searching.

\section{Methodology}

\subsection{Simulation Description}

\begin{figure}
 \includegraphics[width=1\linewidth]{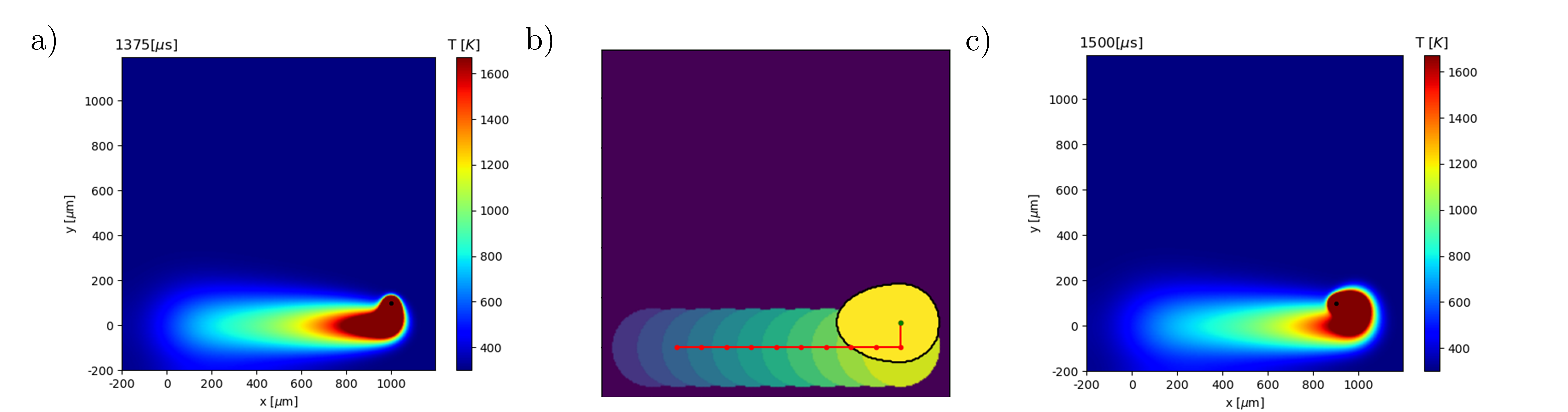}
  \caption{The simulation of the continuum heat distribution using the Repeated Use of Stored Line Solutions Method. (a) The Eagar-Tsai model solves the heat conduction Partial Differential Equation for $\Delta t$, describing the temperature distribution of a moving heat source represented by a Gaussian function on the surface of the domain. (b) To advance the simulation by $\Delta t$ using the RUSLS method, a Gaussian blur is applied to the solution $T(x, y, z, t - \Delta t)$ to represent the heat diffusion far away from the heat source, and then the solution for the ET model for $\Delta t$ and a given set of process parameters is superimposed on the domain to correspond to the laser's position at time $t$. (c) The solution using the RUSLS method in combination with the ET model, at time $t =1500 \mu s $. The maximum scale of the colorbar represents the melting point of the powder.}

\end{figure}


In this work, we consider the heat conduction of a moving heat source in a rectangular domain. To create an efficient simulation that is computationally feasible for Reinforcement Learning, the complex multiscale effects of Powder Bed Fusion are abstracted to the continuum temperature distribution of the material. The process is modeled as the two dimensional conduction associated with a moving heat source on a rectangular plate, with the governing equation given by the following Partial Differential Equation 

\begin{equation}
\frac{\partial T}{\partial t}  = D \nabla^{2} T(\mathbf{x}, t) + \Theta (\mathbf(x), t)
\end{equation}

where $D = \frac{k}{\rho c_p}$ is the thermal diffusivity and $\Theta = \frac{Q}{\rho c_p}$ 
The process is modeled as having uniform thermal properties with parameters listed in Table 1. When Equation 1 is solved using the Green's function for heat conduction in an infinite medium, Equation 2 is produced, describing the temperature field $T (x, t)$. Equation 2 can be decomposed into two separate contributions to the temperature solution, the first term representing the contribution of the heat source $T_Q^{(i)} (x)$  and the second term representing the history of heat diffusion $T_{D}^{i}(x)$. $$T(x, t^{(i)}) = T^{(i)} (x)  = T^{(i)}_Q(x) + T_D^{(i)}(x)$$ The contribution of the heat source can be modelled by the  Eagar-Tsai solution for the conduction, Equation (3), while the diffusion history implementation will be discussed in section 2.3. The method of images is used to implement the boundary conditions.

The general solution, after applying the Green's function solution method is 
\begin{equation}
    T(x,t) = \int_{t_0}^{t} \int_{\Omega} G(\mathbf{x} - \mathbf{x'}, t - \tau)\Theta(\mathbf{x'}, \tau) d \mathbf{x'} d\tau + \int_{\Omega} G(\mathbf{x - x'}, t - t_0)T(\mathbf{x'}, t_0) d \mathbf{x'}
\end{equation}
 
 where $G(x)$ is the Green's function given by
 \begin{equation}
     G(\mathbf{x-x'}, t - \tau) = \frac{1}{[4\pi D(t - \tau)]^{3/2}} \exp{ \left\{ -\frac{||x' - x||^{2} }{4D(t - \tau)} \right\}}
 \end{equation}

The heat source  can be parameterized as a Gaussian distribution moving on the surface of the plate, in Eq (4).

\begin{equation}
    \theta(x, y, z, t)  = \frac{A P}{2 \pi \sigma^2 \rho c_p} \exp \left \{- \frac{(x - Vt)^2 + y^2}{2 \sigma^2} \right \} \delta(z) 
\end{equation}

This leads to the formulation of the Eagar-Tsai model of transient heat conduction, Equation 4, representing the temperature distribution induced by a moving heat source for some $\Delta t$ at velocity $V$ in the $x$-direction.

\begin{equation}
    T_l^{(i)} = \frac{A P}{\rho c_p \sqrt{4\pi^3 D}}\times \int_0^{\Delta t^{(i)}} \frac{\tau^{-1/2}}{\sigma^{2} + 2D\bar{\tau}} \exp \left \{ -\frac{(x+V \bar{\tau}) + y^2}{2 \sigma^2 + 4D\bar{\tau}} - \frac{z^2}{4D\bar{\tau}} \right \} d\bar{\tau}
\end{equation}

The authors of \cite{wolfer2019fast} make use of the method of Repeated Use of Stored Line Solutions to increase the efficiency of the solution process. While fully analytical methods using finite-element methods solve the Eagar-Tsai model for the entire domain, the RUSLS method solves the Eagar-Tsai model for a small line trajectory and reuses the solution to generate the heat distribution in the immediate wake of the laser, after making modifications to account for the geometry of the problem. The solution for the Eagar-Tsai model for a moving point source, from time $t = 0$ to time $t = \Delta t$ is given by Equation (5), which can be appropriately translated and rotated, to represent motion starting in a given location $(x, y)$ and travelling at an angle $\theta$ \cite{eagar1983temperature}. This solution $T_l (i)$ is translated and rotated to the the starting point of the laser (moving from point $(0, 0)$ to point $(x, y)$, and added to the existing temperature distribution, $T’(x,y)$ to form the temperature distribution at time $t$. To continue advancing the laser where there is an existing temperature distribution present, the heat diffusion from time $t$ to time t + $\Delta t$ is modeled to form $T’(x,y)_t$ and the line solution  $T_l (i)$ is once again oriented to the correct location and added to $T’(x,y)$ to form $T(x,y)$. Using these simulations, it is now feasible to iterate over many candidate control policies in a relatively short amount of time, reducing the computational expenses associated with standard Finite Element Analysis methods.

\subsection{Convolution and Boundary Conditions}

Near the boundary of the domain, the Eagar-Tsai model must be modified to produce the appropriate line solutions. If the laser approaches a distance of $4 \sqrt{2 \frac{2 k \Delta t}{\rho c_p}}$from the boundary of the domain, the method of images is used to account for the effect of the boundary on the heat distribution \cite{wolfer2019fast}. Specifically, a virtual heat source is simulated at the same distance away on the other side of the boundary during the computation of the line solution. Therefore, edge solutions and corner solutions can be computed by mirroring the normal solution across the relevant boundaries to account for the interaction that the boundary has with the regular dynamics. This virtual heat source is implemented by modifying the dimension integrals in equation (6). 

\begin{equation}
T_{l}^{(i)} (x, y, z) = \frac{AP}{2 \pi \sigma^2 \rho c_p \sqrt{\pi^3}} \times \int_0^{\Delta t^{(i)}} (4 D \bar{\tau})^{-3/2} \hat{X}(x, \bar{\tau}) \hat{Y} (y, \bar{\tau}) \hat{Z} (z, \bar{\tau}) d \bar{\tau}  
\end{equation}

In order to account for the history of heat diffusion for the existing temperature distribution on the plate, the second term of Eq. 2 is implemented as a convolution operation. Since the Laplacian of a given vector field acts as a local averaging operator, it can be approximated by applying a convolution filter with weights determined by a Gaussian distribution. This operation can be viewed as a Gaussian blur, with strength of the Gaussian blur determined by the thermal properties of the material, the timescale over which diffusion occurs, and the strength of the laser. 

\begin{equation}
  T_D^{(i)} (\mathbf{x}) = \int_{\Omega} G(\mathbf{x} - \mathbf{x'}, \Delta t^{(i)}) T^{(i-1)} (\mathbf{x'}) d \mathbf{x'} 
\end{equation}

\begin{equation}
  T_D^{(i)} (\mathbf{x}) = [G^{(i)} * T^{(i-1)}] 
\end{equation}

As the the convolution filter is carried out by a weighted average of temperature values in an equi-spaced square grid for each pixel in the domain, special considerations must be made near the boundaries where the convolution filter may extend over the boundary of the mesh. In cases where the boundary condition is constrained to be adiabatic, the domain is artificially extended by the size of the convolution filter. The values in this extended section are taken as the mirror image of the temperature values near the boundary. In cases where the boundary condition is constrained to be a specific temperature value, this extended section is populated by the reference temperature value subtracted by the mirror image of the temperature distribution near the boundary.

The meltpool depth is used as a metric for gauging the success of the model, and is calculated by interpolating the temperature field along the y-axis and to find the position with the highest surface temperature, then interpolating along the z-axis to find the point at the temperature below the surface first becomes greater than the melting temperature of the material. This is done by using a root finding algorithm to minimize the distance between the temperature and the melting point of the material based on the current mesh discretization.

\begin{figure}
 \includegraphics[width =  1\linewidth]{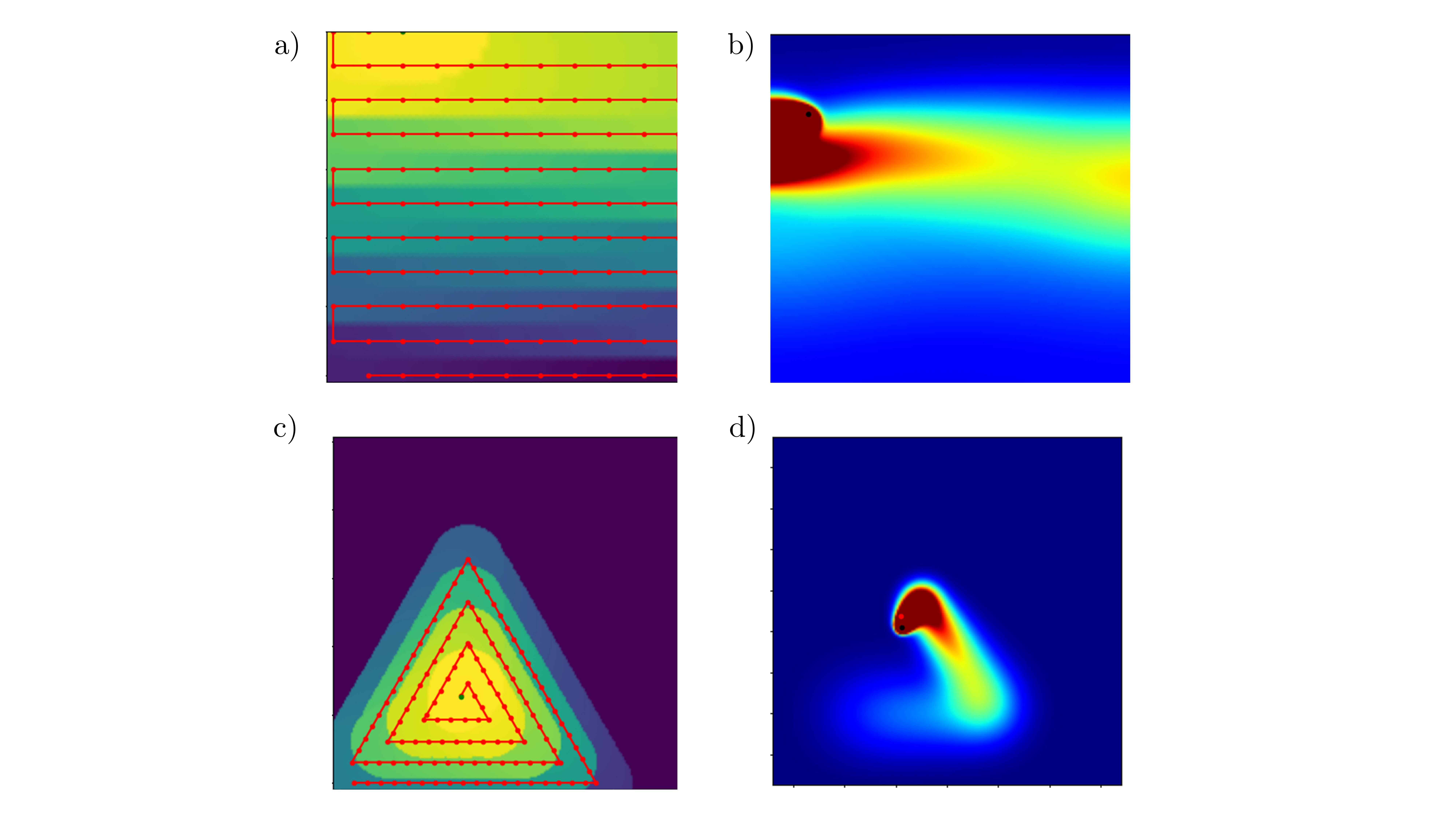}
  \caption{An illustration of the scan paths used to evaluate the performance of the Deep Reinforcement Learning framework. (a) The horizontal cross-hatching scan path. The laser starts at the bottom of the domain, traverses the entire length of the domain, $1.25 mm$ horizontally, advances upwards by $0.125 mm$, and then traverses the domain in the other direction. (b) The heat distribution taken at $ t = 9200 \mu s $ during the trajectory. The boundary near the laser induces overheating, as energy builds up due to the laser changing direction and is unable to dissipate due to the adiabatic boundary condition. (c) The concentric triangular scan path. The laser starts at the bottom of the domain, moves horizontally for 75\% of the length of the domain, then makes a 60$^{\circ}$ pivot and travels 75 \% of the distance it did previously. (d) The heat distribution taken at $ t = 1500 \mu s $ during the concentric triangular trajectory. Overheating is induced when the laser changes direction, and gradually as the residual heat from the tightly packed oscillations begins to build in the center of the domain.}
   
\end{figure}

\begin{figure}
 \includegraphics[width=1\linewidth]{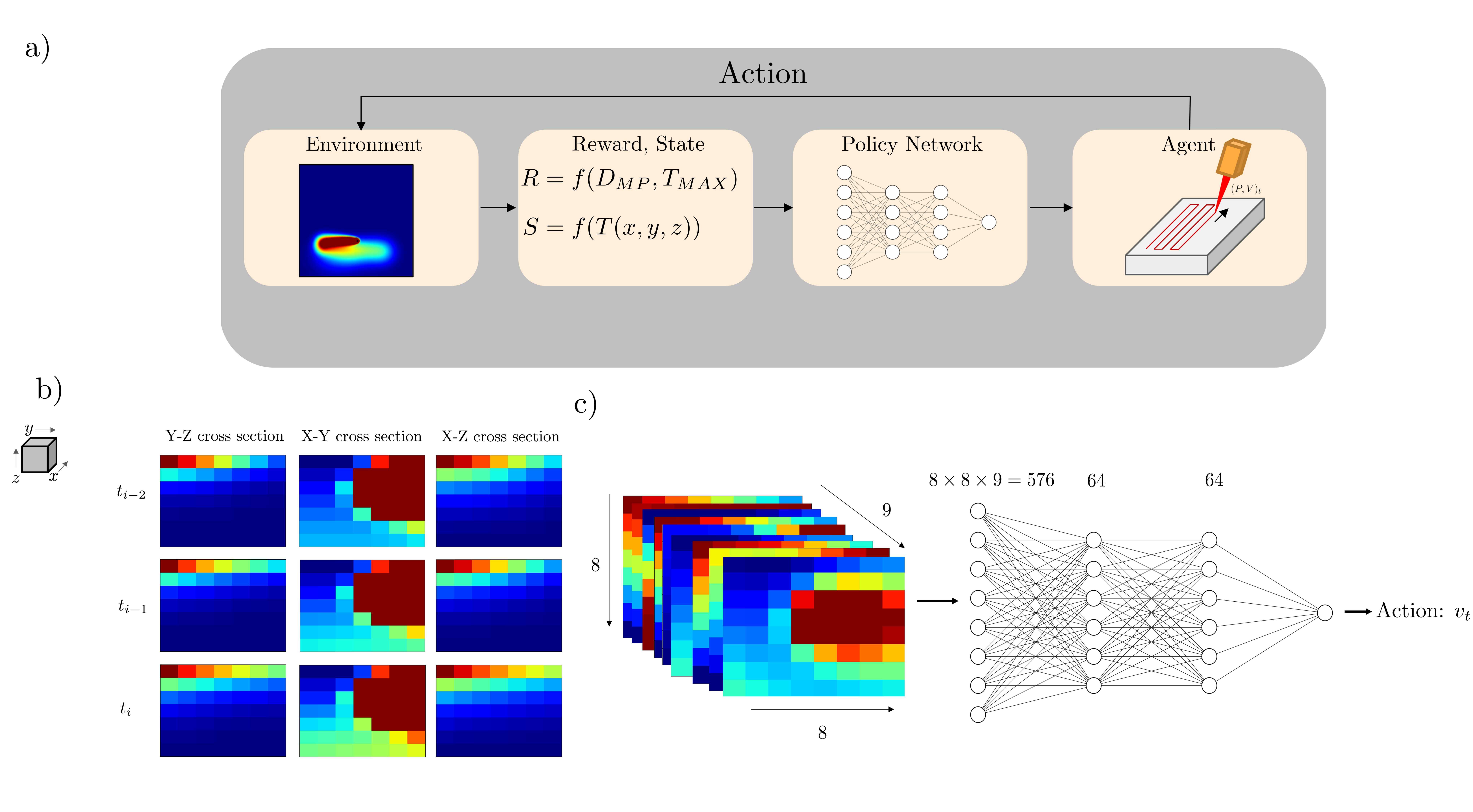}
  \caption{The Deep Reinforcement Learning framework. (a) In Reinforcement Learning, an agent selects an action, based on the current state $s$ and the policy $\pi$ mapping each state to an action. The action is deployed in the environment and the action moves to a different state, influencing the environment and producing a reward based on this influence. The policy is updated based on reward, with the end goal of maximizing the observed reward. (b) The state is represented by cross sections of the domain in the x-y, y-z, and x-z planes near the location of the laser, for the three previous timesteps of the simulation. (c) The policy network is a fully connected neural network that takes in the state space matrix, and predicts an action with the aim of maximizing the expected reward. The Policy Network is implemented as a two layer multi-layer perceptron with hyperbolic tangent activation functions and 64 neurons per hidden layer.}
   
\end{figure}

\subsection{Reinforcement Learning Framework}

In Reinforcement Learning, a policy selects the optimal control action to take, based on input from the environment. This action subsequently influences the environment, and the effect of this influence on the problem that is solved is quantified by a reward. Specifically, a state space S is defined to act as a low-dimensional representation of the current state of the environment, an action space A is defined to represent the potential actions available to the agent, and the reward quantifies the success of the action(s) taken in the previous step at achieving the prescribed goal. 

The reinforcement learning optimization paradigm seeks to maximize the reward collected during an episode. This is done by generating a policy, $\pi$, that will select an action based on the current state, in order to maximize the future expected reward generated from the remainder of the episode. The future expected return of a given state while acting according to policy $\pi$ is termed the \textit{value function}, $V^{\pi} (s)$ and the future expected return of a given state after taking a specific action $a$ and thereafter acting with policy $\pi$ is the \textit{action-value function}, $Q^{\pi} (s, a)$. The policy is iteratively optimized to find the optimal policy $\pi^*$ which maximizes the value of $Q^{\pi} (s,a)$.
\begin{equation}
    Q^{\pi}(s, a)=\underset{s^{\prime} \sim P}{\mathrm{E}}\left[r(s, a)+\gamma \underset{a^{\prime} \sim \pi}{\mathrm{E}}\left[Q^{\pi}\left(s^{\prime}, a^{\prime}\right)\right]\right]
\end{equation}
\begin{equation}
     V^{\pi}(s)=\underset{a \sim \pi \atop s^{\prime} \sim P}{\mathrm{E}}\left[r(s, a)+\gamma V^{\pi}\left(s^{\prime}\right)\right]
\end{equation}
The state space is defined as the observations of the temperature field in specific views and orientations in this formulation. The state space is passed to the policy network as nine 2-D heat maps of the local temperature distribution surrounding the current position of the laser in the domain. A $160 \mu m \times 160 \mu m$ area is defined around the laser, centered on the laser in the x-y cross-sections, and extending downwards from the surface of the domain in the y-z and x-z cross-sections. These three cross-sectional snapshots of the temperature field are concatenated with the two previous sets of snapshots observed during the episode trajectory. Furthermore, the temperature values are whitened, subtracting the mean and dividing by the standard deviation of the state space to approximate a standard normal distribution for the data.

The action space is defined as the process parameter updates that are made to the laser characteristics to alter the behavior of the melting process. Specifically, the velocity of the laser in traveling from one pre-specified point in the trajectory to the next is specified. The actions are also rescaled to lie within [-1, 1], to avoid the issue of vanishing gradients that are common with certain activation functions. 
\begin{equation}
    v_{i+1} = a_{i+1} \times (v_{max} - v_{min}) + v_{min}
\end{equation}

The reward function quantifies the performance of a control policy on an episode. The reward is defined as the absolute error between the target melt depth and the current depth, combined with a regularization term that avoids "cheating" - where the agent exploits the reward function to produce unwanted behavior. The regularization term penalizes the distance between the minimum and maximum melt depth observed during the episode, avoiding abnormal strategies that may cause sudden spikes in the melt depth that are averaged out in the overall reward for an episode.
 
 \begin{equation}
     r_t = \sum_t 1 - \left | \frac{D_{target} - D_{melt}}{0.5 \times D_{target}} \right | - (\max_t{D_{melt}} - \min_t{D_{melt}})
 \end{equation}

\subsection{Proximal Policy Optimization}

In order to optimize the policy network, we use the algorithm Proximal Policy Optimization (PPO) \cite{schulman2017proximal}, a subcategory of Policy Gradient methods. Policy Gradient methods search probabilistically  for optimal policies by gradient ascent. The policy is optimized based on the \textit{advantage function}, $A^{\pi}$ which represents the specific increase or decrease in expected reward generated by pursuing an action in relation to the expected future reward seen on average proceeding from a given state.

\begin{equation}
    A^{\pi} (s, a) = Q^{\pi}(s, a) - V^{\pi}(s,a)
\end{equation}

Proximal Policy Optimization, inspired by a related method, Trust Region Policy Optimization, limits the maximum size of the gradient ascent step based on the relative increase in expected reward that will be observed by the new policy. This particular method was selected because it is much simpler than Trust Region Policy Optimization, and requires less hyper-parameter tuning than comparable Reinforcement Learning methods such as Deep Q-Learning and Actor-Critic optimization. Additionally, it is better suited for continuous control problems, where the potential action space is not discretized into a finite number of separate actions. Policy Gradient methods are episodic, as the policy network is updated after the completion of an episode based on the accumulated reward. In this setting, each episode is defined as the laser completing an entire pass through the entire scan path. A vectorized version of Proximal Policy Optimization is implemented for this project wherein multiple agents are deployed in parallel to collect streams of experience and update the same policy network. Vectorizing the PPO implementation reduces the time taken for the algorithm to gather the requisite amount of experience to learn an optimal policy.

\begin{equation}
    L^{CLIP} (\theta) = \hat{E_t}[min (r_t(\theta)\hat{A_t}, clip(r_t(\theta), 1 - \epsilon, 1 + \epsilon))\hat{A_t}]
\end{equation}

\section{Results and Discussion}\label{sec:others}

We apply the PPO-enabled Deep Reinforcement Learning algorithm described above to optimize the depth of the melt pool formed during a single layer of the manufacturing process. The method is applied to two different trajectories, a horizontal cross-hatching strategy that is commonly used in industrial applications of Laser Powder Bed Fusion (Figure 2a), and a series of concentric triangles that exaggerates the overheating phenomena that occurs in sub-optimal laser trajectories or confined sections of the powder bed (Figure 2c. As the DRL algorithm will discover a policy that changes the process parameters over time, we compare the performance of each control policy to the melt pool depths that result if the process parameters were held constant during the entire melting process.

\begin{figure}[hbt!]
\includegraphics[width=1\linewidth]{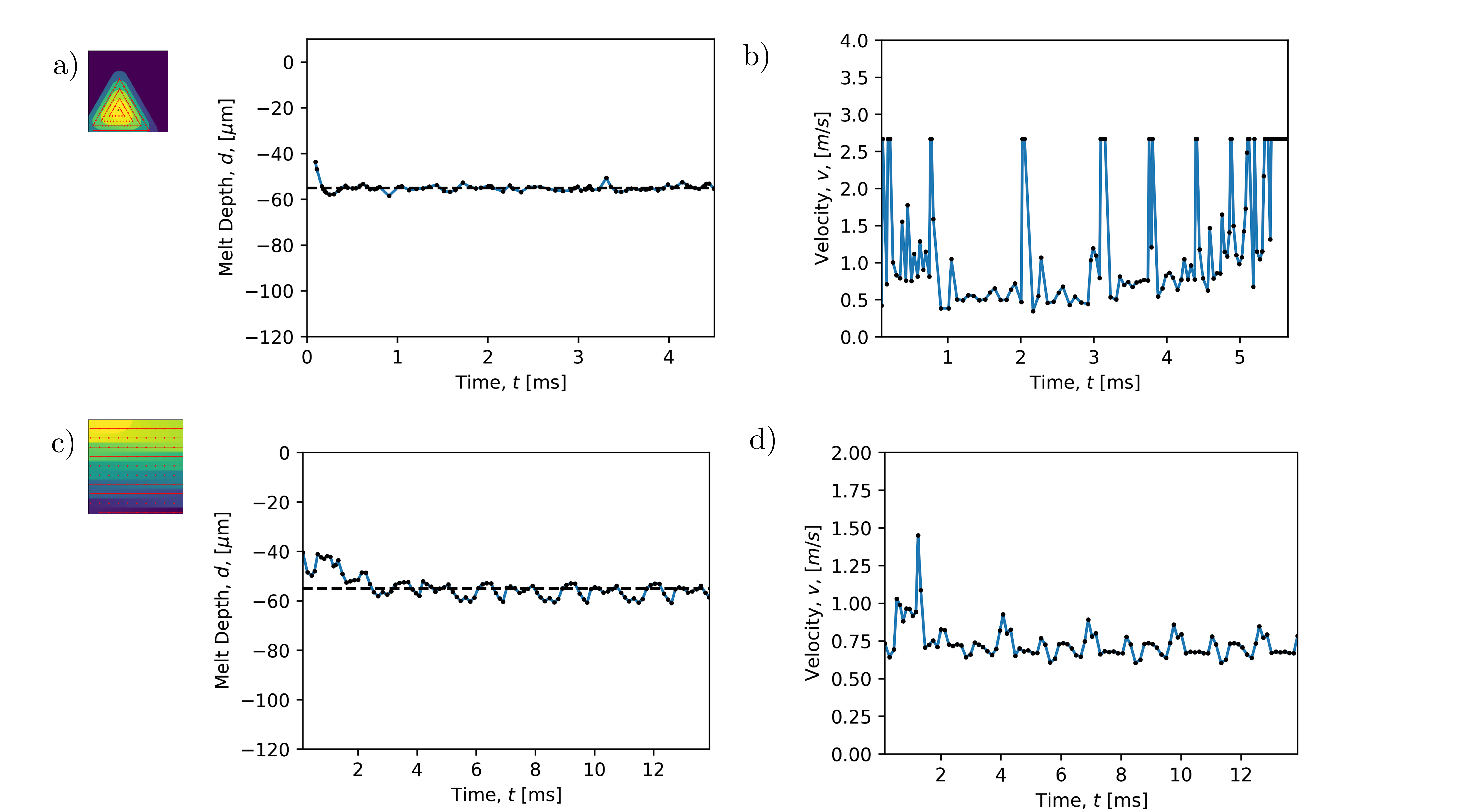}
\caption{(a) The discovered control policy for the horizontal cross-hatching scan path. The velocity increases as the laser reverses directions near the boundaries to reduce the accumulation of thermal energy in these areas. (b) The melt depth achieved for the concentric triangular scan path following the derived control policy. (c) The melt depth achieved for the horizontal cross hatching scan path following the derived control policy. (d) The derived control policy for the concentric triangular path. The velocity increases when the laser changes directions, and the average velocity gradually increases as the laser approaches the center of the scan path.}
\end{figure}

\begin{table}
\caption{Thermal and Process Parameters of the SLM case studied.}
\centering
\begin{tabular}{lll} 
\cmidrule[\heavyrulewidth]{1-3}
Parameters                                      & Values & Units                     \\ 
\cmidrule{1-3}
Absorptivity, $A$                               & 0.3    & -                         \\
Thermal Conductivity, $k$                       & 21.5   & $W/mK$                  \\
Heat Capacity, $c_p$                            & 505    & $J/kg K$                \\ 
\midrule
Mesh discretization, $dx$                       & 20  & $\mu m$  \\
Diameter of laser, $\sigma$                      & 13.75  & $\mu m$                   \\
Initial temperature of powder bed layer, $T_0$  & 300    & $K$                    
\end{tabular}
\end{table}


The Proximal Policy Optimization algorithm is trained for 15,000 episodic updates. The policy network that is used to map states to their corresponding actions consists of two hidden layers, each with 64 neurons and hyperbolic tangent activation functions. The algorithm is trained in parallel on eight environments, and experience from these parallel environments are used to update the model in a synchronous manner. Control actions are taken at predefined intervals of the trajectory of 50 micrometers each, with each of these intervals defined as a single step iteration of the DRL framework. The target melt pool depth that the agent is trained to maintain is $55 \mu m$.The parameters describing the thermal properties of the medium, as well as the dimensions of the laser heat source are listed in Table 1.

Figure 4 displays the discovered control policy for the horizontal cross-hatching trajectory, when strictly controlling the velocity of the laser during the melting process. In the case where the same velocity is used throughout the entire trajectory, there are noticeable peaks in the melt depth at each quarter interval of the trajectory. In these areas, the melt pool depth increases by as much as 20 $\mu m$. The increase in melt depth observed occurs due to the accumulation of energy at the locations at which the laser changes direction, as well as the adiabatic boundary conditions preventing thermal energy from escaping the domain. However, the DRL algorithm successfully optimizes a control policy to limit these effects by modifying the velocity at certain points along the trajectory. As the laser approaches the edges of the domain, the velocity of the laser is increased to decrease the amount of energy transferred to the domain, avoiding spikes in the maximum melt depth due to reduced ability for the heat to diffuse. When compared to the performance of a constant laser velocity, the learned control policy is able to achieve a much smaller variation in the melt pool depth than the constant process parameter scenario. While the melt depth slightly undershoots the target melt depth at points during the melting process, the range that the melt depth occupies is much narrower than the range observed in the uncontrolled scenario. Therefore, assuming that the area of the melt pool can be correlated with the depth of the melt pool at any given point along the trajectory, applying velocity control results in a much more consistent melt pool area that is significantly less susceptible to keyhole formation.

\begin{figure}[hbt!]
\includegraphics[width=1\linewidth]{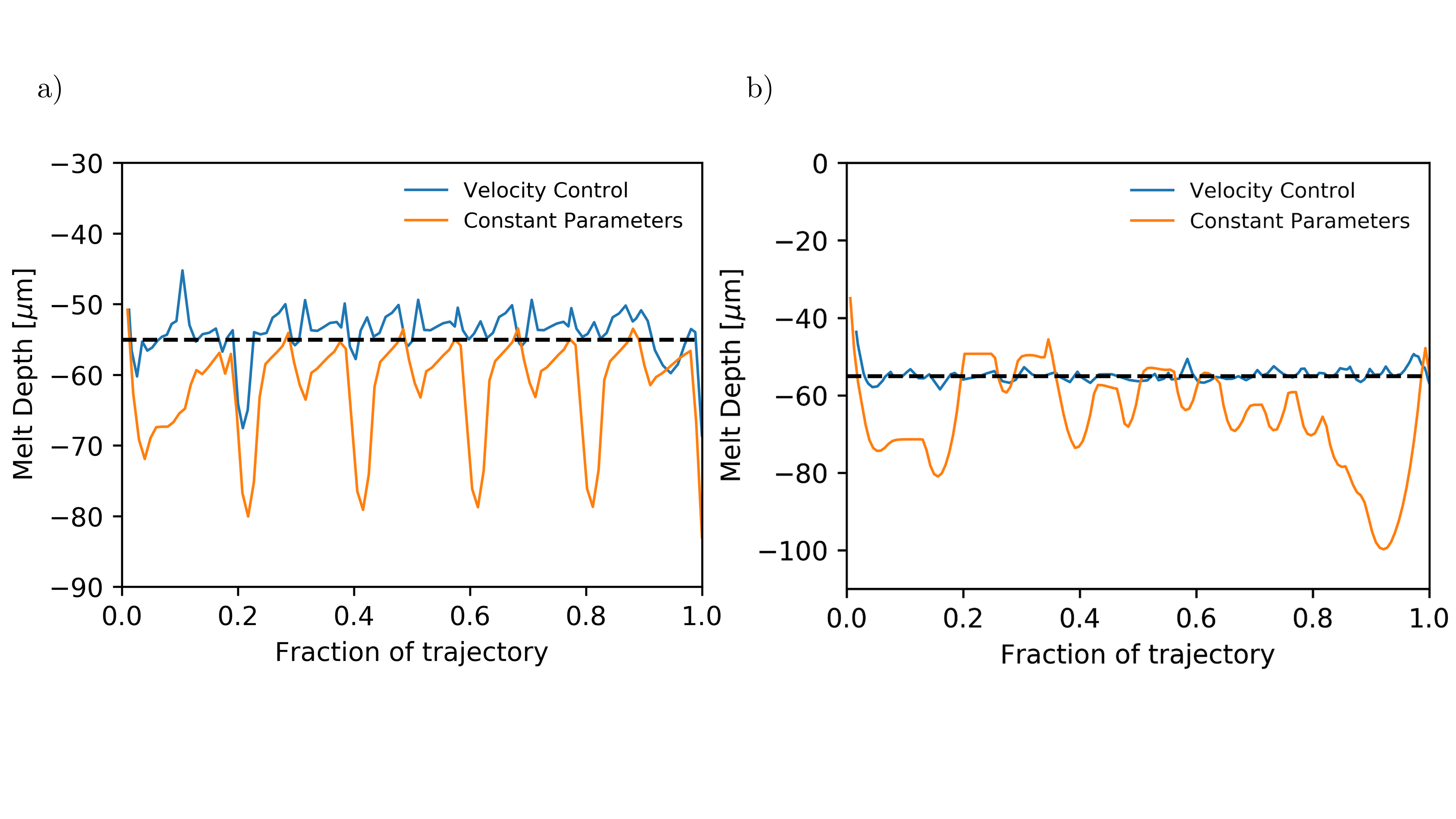}
\caption{(a) The melt depth generated by the control policy compared to the melt depth generated by a constant velocity for the horizontal cross-hatching scan path. The melt pool depth is more stable when compared to a constant velocity applied throughout the entire melting process. (b) The melt depth generated by the control policy compared to the melt depth generated by a constant velocity for the concentric triangular scan path. The melt pool depth is more stable when compared to a constant velocity applied throughout the entire melting process.}

\end{figure}

\begin{figure}[hbt!]
\includegraphics[width=1\linewidth]{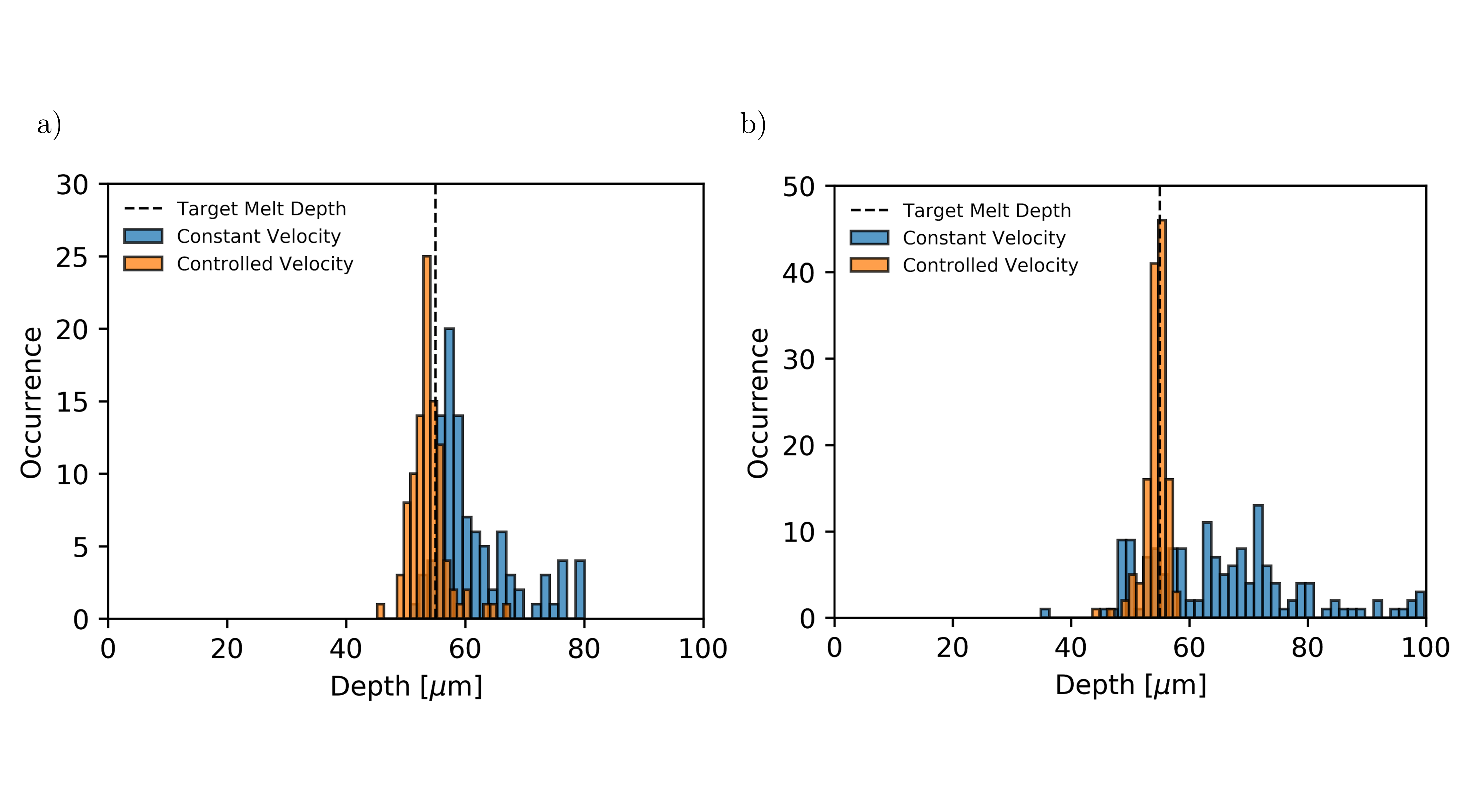}
\caption{(a) A histogram of the melt depth generated by the control policy compared to a histogram the melt depth generated by a constant velocity for the horizontal cross-hatching scan path. The mean of the melt pool depths generated during the melting process is closer to the target melt pool depth, and the standard deviation is smaller. (b) A histogram of the melt depth generated by the control policy compared to the melt depth generated by a constant velocity for the concentric triangular scan path. The mean of the melt pool depths generated during the melting process is closer to the target melt pool depth, and the standard deviation is smaller.}

\end{figure}
Training the algorithm on the concentric triangular trajectory, the control policy is also able to learn a suitable policy by modifying the velocity of the laser as it approaches the center of the domain. In the uncontrolled case, there are large increases in the melt pool depth at each time the laser reverses direction to complete the concentric triangular path. Additionally, towards the end of the trajectory, thermal energy accumulates in the center of the path, due to the overlapping segments of the trajectory and the increased frequency of direction reversal. This thermal energy accumulation can also be seen in the melt pool depth in the last $20\%$ of the trajectory, where there is a sudden increase of $40 \mu m$. When compared to this constant process parameter case, the learned policy is able to avoid the large jump in melt depth that occurs towards the end of the trajectory.  When the laser changes the direction of travel, the velocity increases, similarly to the horizontal cross-hatching scan path. Also, the average velocity of the laser increases as it approaches the center of the scan path, at which point the velocity is held constant at the maximum possible value to reduce overheating. Maximum bounds on velocity can be imposed to avoid the formation of other defects not examined in this work. For instance, spatter can be caused by a high value of jerk in the laser’s motion \cite{Khairallah2016}.

\section{Conclusion}
A Deep Reinforcement Learning method for improving the quality of laser powder bed fusion products is presented. By iteratively optimizing a policy network to maximize the expected reward during the melting process, a control policy to reduce defect formation can be derived through Proximal Policy Optimization. The success of this method is demonstrated through the discovery of effective control policies to reduce the melt pool variability observed in different scan paths in simulation. Specifically, a velocity-based control method is discovered by the  that reduces the degree of overheating experienced due to the geometry of the domain and trajectory of the laser, and decreases the variation of the melt pool depth. By observing the reward generated for specific velocity selections during the melting process, the derived policy intuitively learns to increase the velocity at areas where heat may accumulate. In doing so, the likelihood of phenomena that lead to defect formation is reduced. Additionally, this framework enables closed-loop control, and is generalizable to different trajectories and boundary conditions without requiring domain specific modifications. Extensions of this work can be potentially applied to experimental control, by using deep networks that incorporate physical effects not accounted for in Eagar-Tsai model to create a computationally feasible accurate reinforcement learning environment.

\bibliographystyle{unsrt}  
\bibliography{BibFilesRLAM.bib}  



\end{document}